\documentclass{acm_proc_article-sp}

\usepackage{latexsym,amssymb,amsfonts,amsmath} 
\usepackage[amsmath,thmmarks]{ntheorem}
\usepackage{verbatim,url,graphicx}
\usepackage[latin1]{inputenc}
\usepackage{pstricks,pst-node,epsf,psfrag}
\usepackage{euscript}
\usepackage{multicol}

{
 \theorembodyfont{\upshape} 
 \theoremstyle{nonumberplain} 
 \theoremseparator{.} 
 \newtheorem{xpl}{Running example}} 

\newcommand{\clause}{\mathit{Cl}}
\newcommand{\valid}[0]{\ensuremath{\mathsf{valid}}}
\newcommand{\notassigns}[0]{\ensuremath{\mathsf{not\_assigns}}}
\newcommand{\validacc}[0]{\ensuremath{\mathsf{valid\_acc}}}
\newcommand{\eV}{\EuScript C}
\newcommand{\Nats}[0]{\ensuremath{\mathbb{N}}}
\newcommand{\notinpset}[0]{\ensuremath{\mathsf{not\_in\_pset}}}
\newcommand{\card}[1]{\ensuremath{\textit{card}(#1)}}
\newcommand{\TO}{TO}
\newcommand{\eL}{\EuScript L}

\begin{document}

\title{Graph-based Reduction of Program Verification Conditions}


\numberofauthors{3} 
\author{
\alignauthor
J.-F. Couchot\\
\affaddr{LIFC}\\
\affaddr{University of Franche-Comt\'{e}}\\
\affaddr{F-25030 Besan\c{c}on}\\
\email{couchot@lifc.univ-fcomte.fr}
\alignauthor
A. Giorgetti\\
\affaddr{INRIA CASSIS / LIFC}\\
\affaddr{University of Franche-Comt\'{e}}\\
\affaddr{F-25030 Besan\c{c}on}\\
\email{alain.giorgetti@univ-fcomte.fr}
\alignauthor
N. Stouls\\
       \affaddr{AMAZONES INRIA / CITI}\\
       \affaddr{INSA Lyon}\\
       \affaddr{F-69621 Villeurbanne}\\
       \email{Nicolas.Stouls@insa-lyon.fr}
}
\additionalauthors{}
\date{\today}

\maketitle
\begin{abstract}
Increasing the automaticity of proofs in deductive verification of C
programs is a challenging task. When applied to industrial C programs
known heuristics to generate simpler verification conditions are not
efficient enough. This is mainly due to their size and a high number of
irrelevant hypotheses.

This work presents a strategy to reduce program verification conditions
by selecting their relevant hypotheses. The relevance of a hypothesis
is determined by the combination of a syntactic analysis and two graph
traversals. The first graph is labeled by constants and
the second one by the predicates in the axioms. The approach is
applied on a benchmark arising in industrial program verification.
\end{abstract}




\section{Introduction}
Deductive software verification aims at verifying program properties
with the help of theorem provers. It has gained more interest with the
increased use of software embedded in, for instance, airplanes
commands, cars or smart cards, requiring a high-level of confidence.

In the Hoare logic framework, program properties are expressed by
first-order logical assertions on program variables (preconditions,
postconditions, invariants,~\ldots). The deductive verification method
consists in transforming a program, annotated with sufficiently many
assertions, into so-called \emph{verification conditions} (VCs) that,
when proved, establish that the program satisfies its assertions. In
the KeY system~\cite{KeYBook2007} a special purpose logic and calculus
are used to prove these verification conditions. The drawback of this
approach is has it is specific to a programming language and a target
prover. In contrast, a multi-prover approach is followed by effective
tools such as ESC/Java~\cite{escjava98} for Java programs annotated
using the Java Modeling Language~\cite{Burdy03b},
Boogie~\cite{BarnettLS04} for the C\# programming language, and
Caduceus/Why~\cite{filliatre07cav} for C programs. The latter also
offers Java as input programming language.

A theorem prover is invoked to establish the validity of each
verification condition. One of the challenges in deductive software
verification is to automatically discharge as many verification
conditions as possible. A key issue is that the whole context of a
verification condition is a huge set of axioms modelling not only the
property and the program under verification, but also many features of
the programming language. Simply passing this large context to an
automated prover induces a combinatorial explosion, preventing the
prover from terminating in reasonable time.

Possible solutions to reduce the VC size and complexity are to optimize
the memory model (e.g. by introducing separations of zones of
pointers~\cite{hubert07hav}), to improve the weakest precondition
calculus~\cite{Leino05} and to apply strategies for simplifying
VCs~\cite{Gri00,DenneyFS06,MP06}. This work focuses on the latter. We
suggest heuristics to select axioms to feed automated theorem provers
(ATPs). Instead of blindly invoking ATPs with a large VC, we present
reduction strategies that significantly prune their search space. The
idea behind these strategies is quite natural: an axiom is relevant if
a prover applies it successfully, i.e. without diverging, to establish
the conclusion. Relevance criteria are computed by the combined
traversal of two graphs representing symbol dependencies within axioms.
In the graph of constants edges represent the conjoint presence of two
constants in some ground axiom. In the graph of predicates arcs
represent logical dependencies between predicates occuring in the same
axiom.

In former work~\cite{couchot07FTP}, selection was limited to ground
hypotheses and comparison predicates were not taken into account. This
led to unsatisfactory results, for instance when the conclusion is some
equality between terms. The present work extends selection to context
axioms, comparison predicates and hypotheses with quantifiers.
We propose new heuristics that increase the number of automatically
discharged VCs.

The plan of the article is as follows. Section~\ref{trusted:sec}
presents the industrial C example that has motivated this work. This
case study is a part of the Oslo~\cite{Oslo} secure bootloader
annotated with a safety property. Section~\ref{sec:vcnf} presents the
general structure of a verification condition.
Section~\ref{sec:memorizing} shows how dependencies are stored in
graphs. The selection strategy of hypotheses is presented in
Section~\ref{sub:rel:hyp}. These last two sections are the first
contribution. The second contribution is the implementation of this
strategy as a module of Caduceus/Why~\cite{filliatre07cav}.
Section~\ref{sec:strat} presents experimentation results.
Section~\ref{relatedandconcl} discusses related work, concludes and
presents future work.
\section{Trusted Platform Case Study}
\label{trusted:sec}
Some new challenges for axiom filtering are posed by the context of the
PFC project on Trusted Computing (TC). PFC (meaning trusted platforms
in French) is one of the SYSTEM@TIC Paris Region French cluster
projects.
The main idea of the TC approach is to gain some confidence about
the execution context of a program. This confidence is obtained by
construction, by using a \textit{trusted chain}. A trusted chain is a
chain of executions where each launched program is previously
registered with a tamperproof component, such as the \textit{Trusted
Platform Module} (TPM) hardware chipset.
In this context of TC, we focus on the Oslo~\cite{Oslo} secure loader.
This program is the first step of a trusted chain. It uses some
hardware functionalities of recent CPUs (AMD-V or Intel-TET
  technologies) to initialize the chain and to launch the first
program of the chain.

The main trusted chain properties are temporal, but some recent
works~\cite{JAG_GG06,GS09} propose a method to translate a temporal
property into first-order logic annotations in the code. This method is
systematic and generates a large amount of VCs, including
quantifications and arrays with many links between them. Therefore,
this approach is a good generator for VCs with a medium or low level of
automaticity.
Table~\ref{tab:oslo} gives some factual information about the studied
part of Oslo. The VCs of this benchmark are publicly
available~\cite{osloNS}.

\begin{table}[bht!]

    \begin{tabular}{@{~}l@{~~}l@{~}l@{~}l@{}}
      \multicolumn{4}{l}{\bf Oslo program and specification}\\
      &Code          &$\approx$&\it 1500 lines \\
      &Specification &$\approx$&\it 1500 lines (functional)\\
      &Number of VCs &$\approx$&\it 7300 VCs\\
      \multicolumn{4}{l}{\bf Observed part of Oslo}\\
      &Observed code &$=$&\it       218 lines\\
      &Specification &$\approx$&\it 1400 lines (functional and generated)\\ 
      &Number of VCs &$=$&\it       771 VCs\\
    \end{tabular}

  \caption{Some Metrics about the Oslo Program}
  \label{tab:oslo}
\end{table}

\section{Verification Conditions}\label{sec:vcnf}
\begin{figure}[hbt!]
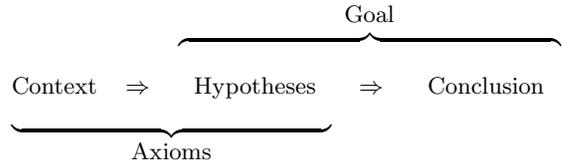

\begin{center}
\begin{tabular}{ccccc}
&  & \multicolumn{3}{c}{Goal} \\
&  & \multicolumn{3}{c}{$\overbrace{\phantom{Accolade horizontale 
vers le bas space}}$} \\
Context & $\Rightarrow$ & Hypotheses & $\Rightarrow$ & Conclusion \\
\multicolumn{3}{c}{$\underbrace{\phantom{Accolade horizontale vers 
haut}}$} \\ %
\multicolumn{3}{c}{Axioms}
\end{tabular}
\end{center}
\caption{Structure of verification conditions \label{fig:vc}}
\end{figure}

The verification conditions (VC) we consider are first order formulae
whose validity implies that a piece of annotated source code satisfies
some property. This section describes the general structure of VCs
generated by Cadu\-ceus/Why. A VC is composed of a \emph{context} and a
\emph{goal}. This structure is illustrated in Fig.~\ref{fig:vc}.

The context depends on the programming language. It is a first-order
axiomatization of the language features used in the program under
verification. Typical features are data types or a memory model,
enriched to allow the specification of, e.g. separated pointer regions.
For instance, a typical VC produced by Caduceus/Why has a context with
more than 80 axioms.

VCs are generated in the input format of many first-order ATPs, among
which Simplify~\cite{simplify05} and SMT solvers~\cite{dMDS07}. The
Simplify automatic prover has a specific input language. SMT solvers
such as Alt-Ergo and Yices have a common input language. Alt-Ergo is
however addressed in the Why input language for more efficiency. For
SMT solvers, the context is presented as a base theory, usually a
combination of equality with uninterpreted function symbols and linear
arithmetic, extended with a large set of specific axioms.

The goal depends on the program and on the property under verification.
When this property is an assertion about a given program control point,
the goal is generated by the weakest precondition (wp) calculus of
Dijkstra~\cite{Dijkstra76} at that control point. The goal is
considered as a \emph{conclusion} implied by \emph{hypotheses} that
encode the program execution up to the control point.

\begin{xpl}
Consider the following function:
\begin{verbatim}
struct p {
 int x;
} p;

struct t {
 struct p v[2];	 
} t;

/*@ requires \valid(a) &&
  @  (\forall int i; 0<=i<=1 => \valid(a->v[i]))
  @ assigns a->v[0].x */
void f(struct t *a) {
 a->v[0].x = 2;
}
\end{verbatim}
The \texttt{requires} annotation specifies a
precondition and the \texttt{assigns} annotation means that function
\texttt{f} modifies no other location than \verb+a->v[0].x+.
The hypotheses of the generated VC are
\begin{displaymath}
\begin{array}{l}
\textsf{valid}($a$),\\
(\forall i:\textsf{int} \,.\, 0 \leq i \leq 1
\Rightarrow \textsf{valid}(a,\textit{shift}(\textit{acc}(m_v,a),i)) 
\ \land  \nonumber \\
\quad \textsf{valid\_acc}(m_{\textit{pPM}})),\\
\textsf{valid\_acc\_range}(m_v,2),\\
\textsf{separa\-tion1\_ran\-ge}(m_v,2),\\
\textsf{va\-lid\_acc}(m_v),\\
r = \textit{acc}(m_v,a),\\
r_0 = \textit{shift}(r,0), \textrm{ and } \\
m_x\_0 = \textit{upd}(m_x,r_0,2).
\end{array}
\end{displaymath}
The conclusion is
\begin{equation*}
\begin{array}{l}
\textsf{not\_assigns}(
 m_x,
 m_x\_0,
 \textit{singleton}(
   \textit{acc}(  
      m_v, 
      a
   )
 )
).
\end{array} 
\end{equation*}
The meaning of these formulae is as follows. $m_{\textit{pPM}}$ is the
pointer ($P$) memory ($M$) for the structures of type \texttt{p}.
\textsf{valid\_\-acc}($m$) means that the memory $m$ is initialized,
i.e. that this memory is accessible from any valid pointer in the
allocation table. The first two hypotheses correspond to the
precondition. In the next two hypotheses the predi\-cates
\textsf{valid\_acc\_range}$(m_v,2)$ and
\textsf{separation1}\_\textsf{range}$(m_v,2)$ respectively mean that
any access to the memory $m_v$ returns an array $t$ such that
poin\-ters $t[0]$ and $t[1]$ are valid and $t[0] \neq t[1]$.  The last
three hypotheses come from a flattening-like decomposition of the
statement \texttt{a->v[0].x = 2} performed by the VC generator. The
function $\textit{shift}(t,i)$ allows access to the index $i$ in the
array $t$. The conclusion translates the \texttt{assigns} annotation
into a relation between two memory values. $m_x$ is the value of memory
$x$ before execution of \verb+f+ and $m_x\_0$ is its value after
execution of \verb+f+. The third parameter is the representation of
\texttt{a->v[0]}. Our preprocessor eliminates the last
three hypotheses and the intermediary constants that they introduce by
considering that the conclusion is
\begin{equation}
\begin{array}{l}
\textsf{not\_assigns}(
 m_x,
 \textit{upd}(m_x,\textit{shift}(\textit{acc}(m_v,a),0),2), \\
 \quad \textit{singleton}(
   \textit{acc}(  
      m_v, 
      a
   )
 )
).
\end{array}  \tag{$C$} 
\end{equation}
\end{xpl}

\section{Graph-Based Dependency}\label{sec:memorizing}
Basically, a conclusion is a propositional combination of potentially
quantified predicates over some terms. Dependencies between axioms and
the conclusion can then arise from terms and predicates. Terms in the
goal may either come from the annotated program (from statements or
assertions) or may result from a weakest precondition calculus applied
to the program and its assertions. The term dependency just transcribes
that parts of the goal (in particular, hypotheses and conclusion) share
common terms. It is presented in Section~\ref{sub:dep:dyn}. Two
predicates are dependent if there is a deductive path leading from one
to the other. The predicate dependency is presented in
Section~\ref{sub:dep:stat}. Finally, Section~\ref{sub:dep:eq} presents
a special dependency analysis for comparison predicates.

\subsection{Term Dependency}\label{sub:dep:dyn}
In order to describe how hypotheses connect terms together and
according to previous work~\cite{couchot07FTP}, an undirected connected
graph $G_c$ is constructed by syntactic ana\-lysis of term occurrences
in each hypothesis of a VC. The graph vertices are labeled with the
constants occurring in the goal and with new constants resulting from
the following flattening-like process. A fresh constant \(f\_i\) where
$i$ is some unique integer is created for each term
\(f(t_1,\ldots,t_n)\) in the goal. There is a graph edge between the
two vertices labeled with the constants $f\_i$ and $c$ when $c$ is
$t_j$ if $t_j$ is a constant and when $c$ is the fresh constant created
for $t_j$ if $t_j$ is a compound term ($1 \leq j \leq n$).

\begin{xpl}
An excerpt of the graph representing the VC presented in
Section~\ref{sec:vcnf} is given in Fig.~\ref{fig:dep_graph_var}.
The vertices \textit{shift}\_6 and \textit{acc}\_7 come from the second 
hypothesis and the other vertices come from the conclusion ($C$).

\begin{figure}[htb!]
\begin{center}
\psfrag{pPM_global}[cc][cc]{$m_{pPM}$}
\psfrag{acc_6}[cc][cc]{\textit{acc}\_6}
\psfrag{shift_6}[cc][cc]{\textit{shift}\_6}
\psfrag{acc_7}[cc][cc]{\textit{acc}\_7}
\psfrag{a}[cc][cc]{$a$}
\psfrag{v_global}[cc][cc]{$m_v$}
\psfrag{acc_5}[cc][cc]{\textit{acc}\_5}
\psfrag{acc_3}[cc][cc]{\textit{acc}\_3}
\psfrag{x_global}[cc][cc]{$m_x$}
\psfrag{shift_2}[cc][cc]{\textit{shift}\_2}
\psfrag{pset_singleton_4}[cc][cc]{\textit{singleton}\_4}
\psfrag{upd_1}[cc][cc]{\textit{upd}\_1}
\includegraphics[width=0.47\textwidth]{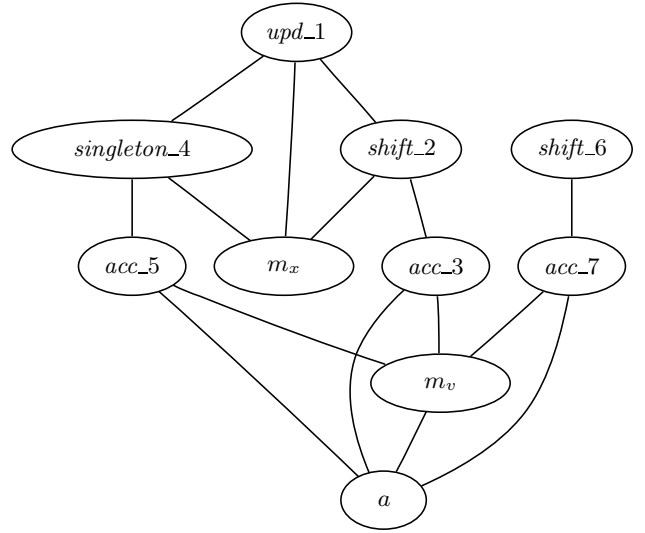}
\end{center}
\caption{Example of Constant Dependency Graph\label{fig:dep_graph_var}}
\end{figure}
\end{xpl}

\subsection{Predicate Dependency}\label{sub:dep:stat}
A weighted directed graph is constructed to represent implication
relations between predicates in an efficient way. Intuitively, each
graph vertex represents a predicate name and an arc from a vertex \(p\)
to a vertex \(q\) means that \(p\) may imply \(q\). What follows are
details on how to compute this graph of predicates, named $G_P$. This
section describes the general approach. The next section adds a special
treatment for comparison predicates.

First, each context axiom is decomposed into a conjunctive normal form
(CNF). It is done in a straightforward way (in contrast to optimised
CNF decomposition~\cite{NW01}): axioms are of short size and their
transformation into CNF does not yield a combinatorial explosion. The
resulting clauses are called \emph{axiom clauses}. Each graph vertex is
labeled with a predicate symbol that appears in at least one literal of
the context. If a predicate $p$ appears negated (as $\neg p$) in an
axiom clause, it is represented by a vertex labeled with
$\overline{p}$. A clause is considered as a set of literals. For each
axiom clause $\clause$ and each pair $(l,l') \in \clause \times
\clause$ of distinct literals in this clause, there is an arc in $G_P$
depending on the polarity of \(l\) and \(l'\). There are three distinct
cases modulo symmetry to consider. They are enumerated in
Table~\ref{table:edge}, where $p$ and $q$ are two distinct predicates.
To reduce the graph size, the contraposite of each implication is not
represented as an arc in the graph but is considered when traversing
it, as detailed in Section~\ref{sub:rel:pred}.

\begin{table}[hbt!]
\begin{center}
\begin{tabular}{@{~~~}c@{~~~}|@{~~~}c@{~~~}}
Value of the $(l,l')$ pair & Arcs \\
\hline
\hline
\( (\neg p,q) \) & $\{p \longrightarrow  q\}$ \\  
\( (p,q) \) & $\{\overline{p} \longrightarrow q \}$ \\ 
\( (\neg p,\neg q)\) & $\{p \longrightarrow \overline{q}\}$  
\end{tabular}
\end{center}
\caption{Translating Pairs of Literals into Arcs.}\label{table:edge}
\end{table}

The intended meaning of an arc weight is that the lower the weight is,
the higher the probability to establish \(q\) from \(p\) is. Therefore,
the arc introduced for the pair \((p,q)\) along Table~\ref{table:edge}
is labeled with the number of predicates minus one (\(card(\clause) -
1\)) in the clause $\clause$ under consideration. For instance, a large
clause with many negative literals, with $\neg p$ among them, and with
many consequents, with $q$ among them, is less useful for a deduction
step leading to \(q\) than the smaller clause \(\{\neg p, q \}\).
Finally, two weighted arcs $p \overset{w_1}{\longrightarrow} q$ and $p
\overset{w_2}{\longrightarrow} q$ are replaced with the weighted arc $p
\overset{min(w_1,w_2)}{-\!\!\!-\!\!\!-\!\!\!-\!\!\!-\!\!\!\longrightarrow} 
q$.

\begin{xpl}
Figure~\ref{fig:dep_graph_axiom} represents the dependency graph
corresponding to the definition of predicates $\valid$, $\notassigns$ and $\validacc$.
It is an excerpt of the graph representing the memory model of
Caduceus/Why.

\begin{figure}
\label{fig:not_assign_dep}
\begin{center}
\psfrag{valid_acc}[cc][cc]{\textsf{valid\_acc}}
\psfrag{valid}[cc][cc]{\textsf{valid}}
\psfrag{valid_acc_C}[cc][cc]{$\overline{\textsf{valid\_acc}}$}
\psfrag{valid_C}[cc][cc]{$\overline{\textsf{valid}}$}
\psfrag{not_assigns}[cc][cc]{\textsf{not\_assigns}}
\psfrag{not_assigns_C}[cc][cc]{$\overline{\textsf{not\_assigns}}$}
\psfrag{not_in_pset}[cc][cc]{\textsf{not\_in\_pset}}
\psfrag{not_in_pset_C}[cc][cc]{$\overline{\textsf{not\_in\_pset}}$}

\includegraphics[width=0.47\textwidth]{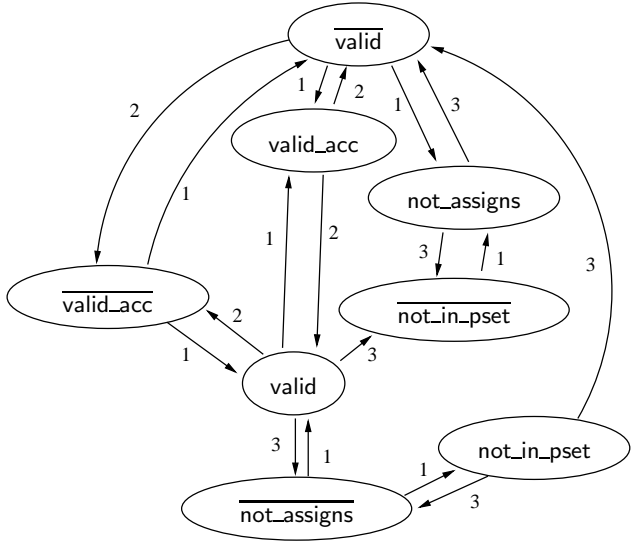}
\end{center}
\caption{Example of Predicate Dependency Graph
\label{fig:dep_graph_axiom}}
\end{figure}
\end{xpl}

\subsection{Handling Comparison Predicates}\label{sub:dep:eq}
In a former work~\cite{couchot07FTP}, equalities and inequalities were
ignored when memorizing predicate dependencies. This leads to
unsatisfactory results when (in)equality is central for deduction, e.g.
when the conclusion is some equality between terms. If we handle
equality as the other predicates, the process of
Section~\ref{sub:dep:stat} connects too many vertices with the vertex
labeled $=$. We have experienced that this reduction of the graph
diameter has a negative impact on the quality of selection.

More generally the present section suggests a special construction of
graph vertices and edges for comparison predicates. A comparison
predicate is an equality $=$, an inequality $\neq$, a (reflexive) order
relation ($\leq$ or $\geq$) or an irreflexive pre-order ($>$ or $<$).
The keys of this construction are the support of types and the
exploitation of some causalities between comparison predicates.

\subsubsection{Typed comparisons}
Each comparison predicate $\circ$ is written $\circ_t$ where $\circ$ is
$=$, $\neq$, $\le$, $<$, $\ge$ or $>$ and $t$ is the type of the
$\circ$ operands. For simplicity, the focus is on
 the types $t$ where $\le_t$ and $\ge_t$ are total orders, $>_t$ and
 $<_t$ are their respective reverse orders, and $\le_t$ is the union of
 $<_t$ and $=_t$. A typical example is the type \textsf{int} of
 integers.

Each comparison $t_1 \circ_t t_2$ present in at least one axiom
is represented by two nodes respectively labeled with $\circ_t$ and
$\overline{\circ_t}$, where $\overline{=_t}$, $\overline{\neq_t}$,
$\overline{\le_t}$, $\overline{<_t}$, $\overline{\ge_t}$, and
$\overline{>_t}$ respectively are $\neq_t$, $=_t$, $>_t$, $\ge_t$,
$<_t$, and $\le_t$. For instance, the two nodes $\le_{\textsf{int}}$
and $>_{\textsf{int}}$ represent a total order on integers and its
negation. These labels are called the \emph{typed} comparison
predicates.

Apart from this difference in the definition of $\overline{\circ_t}$,
the arcs connected to typed comparison predicates are constructed
following the general rules described in Table~\ref{table:edge}.

\subsubsection{Causalities between comparison predicates}
Verification conditions are expressed as SMT problems in AUFLIA
logics~\cite{RanTin-SMTLIB}. Since the comparison predicates between
integers are interpreted in AUFLIA, no context axiom contributes to
their definition. Figure~\ref{comp:ax:fig} suggests such a list of
axioms. To lighten the figure, the predicates are not
indexed with \textsf{int}.

Adding these axioms to the context would be counterproductive. We
propose instead to analyze them to enrich the predicate graph \emph{as
if} they were in the context. Since the algorithm of axiom selection
does not take loops into account, the sole arcs of interest in the
predicate graph are between distinct nodes. It is then impossible to
proceed so on internal properties like reflexivity, irreflexivity,
symmetry or transitivity. This is the reason why
Figure~\ref{comp:ax:fig} is limited to axioms between distinct
predicates. The symmetric axioms where $\le$ and $<$ respectively
replace $\ge$ and $>$ are also treated but are not reproduced. The arcs
resulting from the application of the rules of Table~\ref{table:edge}
to those ten axioms are added to the graph of predicates.

 \begin{figure}
\begin{eqnarray}
x \le y \land y \le x & \Rightarrow & x = y \\
x = y & \Rightarrow & x \ge y \\
x = y & \Rightarrow & y \ge x \\
x > y & \Rightarrow & x \ge y \\
x \ge y & \Rightarrow & x > y \lor x = y
\end{eqnarray}
\caption{Some Axioms Relating Comparison Predicates
\label{comp:ax:fig}}
\end{figure}

\section{Axiom Selection}\label{sub:rel:hyp}
Relevant axioms remain to be selected. Intuitively, an axiom is
relevant with respect to a conclusion if a proof that needs this axiom
can be found. Variables and predicates included in a relevant axiom are
also called relevant.

Section~\ref{sub:rel:var} shows how to select relevant constants in,
Section~\ref{sub:rel:pred} how to select relevant predicates and
Section~\ref{sub:hyp:sel} how to combine these results to select
relevant axioms. A selection strategy is presented as an algorithm in
Section~\ref{select:algo:ssec}.

\subsection{Relevant Constants}\label{sub:rel:var}
A node in the graph of constants $G_c$ is identified with its 
labeling constant. Let $n$ be the diameter of the graph of constants 
$G_c$. Starting from the set \({\eV}_0\) of constants in the 
conclusion, a breadth-first search algorithm computes the sets 
\({\eV}_i\) of constants in $G_c$ that are reachable from \({\eV}_0\) 
with at most $i$ steps ($1 \le i \le n$). Finally, unreachable 
constants are added to the limit of the sequence \( \big( 
{\eV}_n\big)_{n \in \Nats} \) for completeness. Let ${\eV}_{\infty}$ 
be the resulting set.

To introduce more granularity in the computation of reachable
constants, we propose as a heuristic to insert nodes that are linked several 
times before nodes that are just linked once. Semantically it gives 
priority to constants which are closer to the conclusion. Notice 
that, in this case, the index $i$ of \({\eV}_i\) does not correspond 
to a path length anymore.

\begin{xpl}
The sequence of reachable constants sets associated to the
graph in Fig.~\ref{fig:dep_graph_var} is:
\[ 
  \begin{array}{lcl}
    {\eV}_{0} & = & 
    \{ m_x, m_v, a \}, \\
    {\eV}_{1} & = &  
    {\eV}_{0} \cup \{\textit{acc}\_3, \textit{acc}\_5,
    \textit{acc}\_7\}, \\
    {\eV}_{2} & = &  
    {\eV}_{1} \cup \{ \textit{singleton}\_4,
    \textit{shift}\_2 \},\\ 
    {\eV}_{3} & = &  
    {\eV}_{2} \cup \{\textit{shift}\_6 \},\\
    {\eV}_{4} & = &   
    {\eV}_{3} \cup \{\textit{upd}\_1 \} \textrm{ and} \\
    {\eV}_{\infty} & = & {{\eV}_{4}.}
  \end{array}
\]
\end{xpl}

\subsection{Relevant Predicates}\label{sub:rel:pred}
A predicate $p$ is identified with the vertex labeled $p$ and its
negation with the vertex labeled $\overline{p}$
 in the graph of predicates $G_P$. A predicate 
symbol \(p\) is relevant w.r.t. a predicate symbol \(q\) if there is 
a path from \(p\) to \(q\) in $G_P$, or dually from \(\overline{q}\) 
to \(\overline{p}\). 
Intuitively, the weaker the path weight is, the 
higher the probability of~\(p\) to establish~\(q\) is. Relevant 
predicates extracted from~$G_P$ are stored into
an increasing sequence $(\eL_n)_{n \in \Nats}$ of sets.
The natural number $n$ is the maximal weight of paths considered in
the graph of predicates.

We now present how $\eL_n$ is computed.
The conclusion is assumed to be a single clause.
$\eL_0$ gathers the predicates from the conclusion.
For each predicate symbol $p$ that is not in $\eL_0$, a graph traversal
computes the paths with the minimal  weight $w$ from $p$ to some
predicate in \(\eL_0\).

Furthermore, contraposition of each implication is considered: 
let  $p_1$ and $p_2$ be two node labels, corresponding either
to a positive or a negative literal.
If the arc $p_1 \overset{w}{\longrightarrow} p_2$ is taken into account, 
its couterpart $\overline{p_2} \longrightarrow \overline{p_1}$ is too,
with the convention that $\overline{\overline{p}}$ is $p$.
Let $n$ be the minimal distance from $\eL_0$ to the deepest 
reachable predicate. For $1 \le i \le n$, $\eL_i$ is the set of 
vertices of $G_P$ whose distance to $\eL_0$ is less than or equal 
to $i$. $\eL_{\infty}$ is the
limit $\bigcup_{i \geq 0} \eL_i$ 
augmented with the vertices from which $\eL_0$ is not reachable.

\begin{xpl}
From the predicate graph of the running example, depicted in
Fig.~\ref{fig:not_assign_dep} without the comparison predicates for
lack of space, the first five sets of reachable predicates are
\[
\begin{array}{llllll}
{\eL}_0 & = &\{\notassigns\},\\
{\eL}_1 & = & {\eL}_0 \cup \{\overline{\valid},
\overline{\notinpset}, = \}, \\
{\eL}_2 & = & {\eL}_1 \cup \{
  \overline{<_{\textsf{int}}},
  \overline{\validacc},
  \overline{\leq_{\textsf{int}}}
  \}, \\ 
{\eL}_3 & = & {\eL}_2 \cup \{
  \validacc,
  \overline{>_{\textsf{int}}},
  \overline{\neq_{\textsf{int}}},
  \overline{\geq_{\textsf{int}}}
  \} \textrm{ and }\\
{\eL}_4 & = & {\eL}_3 \cup \{
  \overline{=},
  \notinpset,
  \valid,   
  \leq_{\textsf{int}},
  \overline{=_{\textsf{int}}}
\}.
\end{array}
\]
\end{xpl}

\subsection{Selection of Relevant Axioms}\label{sub:hyp:sel}
In this section, we present the main principles of the axiom selection
combining predicate and constant selection. A first part describes
hypothesis selection and a second one extends the approach to axioms
from the context.
 
Let $(\eL_n)_{n \in \Nats}$ and \( \big( {\eV}_n\big)_{n \in \Nats}\)
respectively be the sequences of relevant predicate and constant sets.
Let $i$ be a counter which represents the depth of predicate selection.
Similarly, let $j$ be a counter corresponding to the depth of constant
selection.

\subsubsection{Hypothesis Selection}
Let \(\clause\) be a clause from a hypothesis.
Let \(V\) be the set of constants of \(\clause\) augmented with 
constants resulting from flattening (see Section~\ref{sub:dep:dyn}). 
Let $P$ be the set of predicates of \(\clause\).
The clause \(\clause\) should be selected if 
it includes constants or predicates that are relevant according 
to the conclusion.
Different criteria can be used to verify this 
according to its sets \(P\) and \(V\).
Possible choices are, in increasing order of selectivity 
\begin{enumerate}
\item  
 the clause includes at least one relevant constant or one
 relevant predicate: 
 
\centerline{$V \cap  {\eV}_{j}  \neq \emptyset ~\lor~ P \cap {\eL}_i  \neq \emptyset $}

\item \label{item:stat} 
 the clause includes more than a threshold $t_v$ of relevant
 constants or  more than a threshold $t_p$ of 
 relevant predicates: \vspace*{3pt}

\centerline{$  {\card{V \cap {\eV}_{j}}/\card{{\eV}_{j}} \ge t_v} ~\lor 
     {\card{P \cap {\eL}_i}/\card{{\eL}_i} \ge t_p}$}

\item \label{item:include} 
  all the clause constants and clause predicates are relevant:

\centerline{$V \subseteq {\eV}_{j} ~\land~ P \subseteq {\eL}_i$}

\end{enumerate}

Our experiments on these criteria have shown that a too weak criterion
does not accomplish what it is designed for: too many clauses  are
selected for few iterations, making the prover quickly diverge. Thus,
we only consider the strongest criterion~(3).

We have also often observed the case where only a conjunctive part of a
universally quantified hypothesis is relevant. In that case, we split
the conjunctive hypothesis into its parts and the filtering criterion
is applied to the resulting predicates. A particular case is considered
if a whole splittable hypothesis is relevant according to the
criterion. Indeed, we then consider the original formula, in order to
preserve its structure, which can be exploited by provers.

\subsubsection{Context Axioms}
Consider now the case of selecting relevant axioms from the context.
Intuitively, an axiom of the context has to be selected if 
one of the predicate relations it defines is relevant for one
hypothesis, i.e. the corresponding arc is used in the computation of $\eL_i$. 
Practically, for each arc that is passed through while generating
$\eL_i$, we keep all the axioms of the context that have generated 
this arc.

\subsection{Selection Strategy}
\label{select:algo:ssec}

\begin{figure}\footnotesize
  \[
  \begin{array}{|l|}
    \hline
    \textnormal{Parameters : }\textit{VC, Prover, \TO}\\[2pt]
    \left\lfloor
      \begin{array}{l}
        \textit{// Prover call without VC reduction}\\
        \textit{Res}:=\textit{Prover}(\textit{VC}, \TO)\\ 
        \textit{\textbf{if}} ~ \textit{Res} = \textit{timeout} ~ \textit{\textbf{then}} \\
        \left\lfloor
          \begin{array}{l}            
            i_\textit{max} \!:=\! 1+\textit{Min depth giving reachable preds (VC)}\!\!\!\!\\
            j_\textit{max} \!:=\! 1+\textit{Min depth giving reachable vars (VC)}\!\!\!\!\\
            \\
            i:=0 ;\\
            j:=0 ;\\
            \textit{\textbf{While}} ~ \textit{Res} \!\neq\! \textit{unsat}  
            ~\land~ i\le i_\textit{max} \textit{\textbf{ do}}\\
            \left\lfloor
              \begin{array}{l}
                \textit{// Prover call after VC reduction}\\
                \textit{Res}:=\textit{Prover}(\textit{selection$(\textit{VC},i,j)$},
                \TO)\!\!\!\!\!\!\!\!\!\!\!\!\\ 
                j:=j+1 ;\\
                \textit{\textbf{if }}j>j_\textit{max}\textit{\textbf{ then }} \\
                \left\lfloor
                  \begin{array}{l}
                    i:=i+1 ; \\
                    j:=0 ;
                  \end{array}\right.
              \end{array}\right.\\
          \end{array}\right.\\
        \textit{return }Res;
      \end{array}\right.\\[2pt]
    \hline
  \end{array}
  \]
  \caption{General Algorithm Discharging a VC with Axiom Selection}
  \label{fig:algorithm}
\end{figure}

The selection strategy experimented in this work is described in
Fig.~\ref{fig:algorithm}. The algorithm takes three parameters in
input:
\begin{itemize}
  \item a \textit{VC} whose satisfiability has to be checked, 
  \item a satisfiability solver \textit{Prover}, and
  \item a maximal amount of time $\TO$ given by the user to the
  satisfiability solver to discharge the VC.
\end{itemize}

The algorithm starts with a first attempt to discharge the VC 
without axiom selection. It stops if this first 
result is unsatisfiable or satisfiable. Notice that in the latter
case, removing axioms cannot modify the result.
Otherwise, \textit{Prover} is called following an incremental
constant-first selection.

The two natural numbers $i_{\textit{max}}$ and $j_{\textit{max}}$ are depth 
bounds for ${\eL}_{i}$ and ${\eV}_{j}$ computed during predicate graph
and constant graph traversals. Since we want to reach ${\eL}_{\infty}$ and $\eV_{\infty}$,  
$i_{\textit{max}}$ and $j_{\textit{max}}$ are initially computed by the tool as
one plus the minimal depth to obtain all reachable predicates and
constants. This is interpreted by the tool as the $\infty$ depth, according 
to Sec.~\ref{sub:rel:pred} and~\ref{sub:rel:var} (all predicates and
constants of the graphs).

The \textit{selection} function implements the selection of axioms (from
context or hypotheses) according to the strongest criterion~(3).
Discharging the resulting reduced VC into a prover
can yield three outcomes: satisfiable, unsatisfiable or timeout.

\begin{enumerate}
\item If the formula is declared to be unsatisfiable, the procedure
  ends. Adding more axioms cannot make the problem
  satisfiable. 
\vspace*{5pt}

\item If the formula is declared to be  satisfiable, we may have
omitted some axioms; we are then left to increment either $i$ or
$j$, i.e. to enlarge either the set of selected predicates 
or the set of selected constants.

However, allowing predicates has a more critical impact than allowing
new constants, since constants do not appear in context axioms.
Therefore we recommend to first increment $j$, increasing $\eV_j$ until
eventually $\eV_{\infty}$, before considering incrementing $i$. In this
later case, $j$ resets to $0$. 
\vspace*{5pt}

\item If the formula is not discharged in less than a given time,
  after having iteratively incremented $i$ and $j$, then the
  algorithm terminates.
\end{enumerate}

\section{Experiments}\label{sec:strat}
The proposed approach is included in a global context of annotated C
program certification. A separation analysis that strongly simplifies
the verification conditions generated by a weakest precondition
calculus, and thus greatly helps to prove programs with pointers has
been proposed by T.~Hubert and C.~March{\'e}~\cite{hubert07hav}. Their
approach is supported by the Why tool. The pruning heuristics presented
here are developed as a post-process of this tool.

Section~\ref{sub:exp:method} gives some implementation and
experimentation details. Section~\ref{sub:oslo:res} presents
experimental results on an industrial case study for trusted computing.
This case study raises new challenges associated to the certification
of C programs annotated with a temporal logic formula.
Section~\ref{sub:exp:whybenchs} finally gives results obtained on a
public benchmark.

\subsection{Methodology}\label{sub:exp:method}
All the strategies presented in this work are implemented in OCaml as
modules in the Why~\cite{filliatre07cav} tool in less than 1700 lines
of code. Since these criteria are heuristics, their use is optional,
and Why has command line arguments which allow a user to enable or
disable their use. In the current version, several others heuristics
have been developed, which are not considered because their impact on
the performance of Why seems to be less obvious. In order to use the
presented algorithms, the arguments to include in the Why call are:

\centerline{
  \scriptsize
  \texttt{-\,-prune-with-comp -\,-prune-context  -\,-prune-coarse-pred-comp}}
\centerline{
  \scriptsize
  \texttt{-\,-prune-vars-filter CNF}}

The first parameter includes comparison predicates in the predicate
dependency graph. The second one requires filtering not only hypotheses
but also axioms from the context. The third one requires to ignore arc
weights. This option gives better execution times on the Oslo
benchmark. Finally, the fourth argument requires for rewriting
hypotheses into CNF before filtering.

The whole experiment is done on an Intel T8300@2.4GHz with 4Gb of 
memory, under a x86\_64 Ubuntu Linux.

\subsection{Results of Oslo Verification}\label{sub:oslo:res}
First of all, among the 771 generated VCs, 741  are directly 
discharged, without any axiom selection. Next, the approach
developed in~\cite{couchot07FTP} increases the result to
752~VCs.

Among the remaining unproved VCs, some rely on quantified hypotheses
and others need comparison predicates that are not handled 
in the previous work~\cite{couchot07FTP}.
They have motivated the present extensions, namely
CNF reduction, comparison handling and context reduction. 
Thanks to these improvements, 10 more VCs are automatically
proved by using the algorithm described in 
Fig.~\ref{fig:algorithm} with the three provers Simplify,
Alt-Ergo 0.8 and Yices 1.0.20 with a timeout $\TO$ of 10 seconds.

\begin{figure}[ht]
  \begin{center}
    \includegraphics[width=0.47\textwidth]{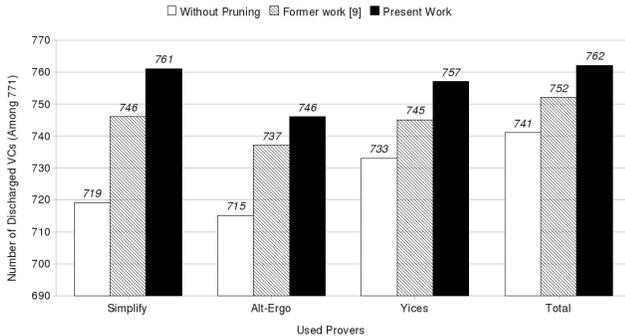}\\   
  \end{center}
  \caption{Result Comparison on Oslo Benchmark (771 VCs)}
  \label{fig:ComparisonDiagramme}
\end{figure}

The $i_\textit{max}$ and $j_\textit{max}$ limits depend on the VCs.
Their observed values do not go beyond $i_\textit{max}=6$ and
$j_\textit{max}=7$. These limits express the number of versions in
which the VCs have been cut. If edge weights are considered, then
$i_\textit{max}$ grows up to $i_\textit{max}=18$ and the execution time
is twice as long.
Figure~\ref{fig:ComparisonDiagramme} sums up these results.

\subsection{Public Why Benchmark}\label{sub:exp:whybenchs}
Our approach is developed in the Why tool, which translates Why syntax
into the input syntax of several proof assistants (Coq, HOL 4, HOL
Light, Isabelle/HOL, Mizar, PVS) and automated theorem provers
(Alt-Ergo, CVC3, Simplify, Yices, Z3). This section shows some
experimental results on the Why public
benchmark\footnote{\url{http://proval.lri.fr/why-benchmarks/}}.

The Why benchmark is a public collection of VCs generated by Caduceus
or Krakatoa. These tools generate VCs respectively from C and Java
programs, according to CSL and JML specifications. Hence, it partially
matches to our requirements, since our work is focusing on the
verification of VCs generated by these tools. The only limitation is
that our method is focusing on VCs with a large amount of hypotheses,
in contrast to the ones presented in this benchmark.

This benchmark is provided in two versions corresponding to two
different pre-processes. Our results are similar with both versions.
Alt-Ergo discharges 1260 VCs directly and 1297 VCs with axiom
selection, while axiom selection adds 3 VCs to the 1310 VCs 
directly discharged by Simplify.

\section{Related Work and Conclusion}\label{relatedandconcl}
We have presented a new strategy 
to select relevant hypotheses in formulae coming from program 
verification. To do so, we have combined two separate dependency 
analyses based on graph computation and graph traversal. Moreover, we 
have given some heuristics to analyse the graphs with a sufficient 
granularity. Finally we have shown the relevance of this approach 
with a benchmark issued from a real industrial code.

Strategies to simplify the prover's task have been widely studied since
automated provers exist~\cite{WosRC65}, mainly to propose more
efficient deductive systems~\cite{WosRC65,WosP99,Wos01}. The KeY
deductive system~\cite{KeYBook2007} is an extreme case. It is composed
of a large list of special purpose rules dedicated to
JML-annotated JavaCard programs. These rules make unnecessary an
explicit axiomatization of  data types, memory model, and program
execution. Priorities between deduction rules help in effective 
reasoning. Beyond this, choosing rules in that framework requires
as much effort as choosing axioms when targeting general purpose
theorem provers. 

The present work can be compared with the set of support (sos)
selection strategy~\cite{WosRC65,PlaistedY03}. This approach starts
with asking the user to provide an initial sos: it is classically the
 conclusion negation and a subset of hypotheses. It is then restricted
 to only apply inferences with at least one clause in the sos,
consequences being added next into the sos. Our work can also be viewed
as an automatic  guess of the initial sos guided by the formula to
prove. In this sense, it is close to~\cite{MP06} where initial relevant
clauses are selected according to syntactical criteria, i.e. counting
matching rates between symbols of any clause and symbols of clauses
issued from the conclusion. By considering syntactical filtering on
clauses issued from axioms and hypotheses, this latter work does not
consider the relation between hypotheses, formalized by axioms of  the
theory: it  provides a reduced forward proof. In contrast, by analyzing
dependency graphs, we simulate natural deduction and are not far from
backward proof search. By focusing on the predicative part of the
verification condition, our objectives are dual to those developed
in~\cite{Gri00}: this work concerns boolean verification conditions
with any boolean structure whereas we treat predicative formulae whose
symbols are axiomatized in a quantified theory. Even in a large set of
context axioms, most of the time, each verification condition only
requires a tiny portion of this context. In~\cite{RF97theorem,DR06} a
strategy to select relevant context axioms is presented, but it needs a
preliminary manual task classifying axioms. Our predicate graph
computation makes this  axiom classification automatic. Recent advances
have been made in the direction of semantic selection of
axioms~\cite{SutcliffeP07,pudlak07}. Briefly speaking, at each
iteration, the selection of each axiom depends on the fact whether a
computed valuation is a model of the axiom or not. By comparison, our
syntactical axiom selection is more efficient, indeed linear in the
size of the input formula.

In a near future we plan to apply the strategy to other case studies.
We also plan to investigate the impact on execution time of various
strategies discharging the same list of verification conditions. We
want to confirm or infirm with other benchmarks that weighting
predicate dependencies with a formula length has no positive impact on
automaticity but has a significant negative impact on the execution
time. We also plan to integrate selection strategies in the Why tool or
in a target automated theorem prover.

\section{Acknowledgments}
This work is partially funded by the French Ministry of Research,
thanks to the CAT (C Analysis Toolbox) RNTL (Reseau National des
Technologies Logicielles), by the SYSTEM@TIC Paris Region French
cluster, thanks to the PFC project (Plateforme de Confiance, trusted
platforms), and by the INRIA, thanks to the CASSIS project and the
CeProMi ARC. The authors also want to thank Christophe Ringeissen and
the four anonymous referees for their insightful comments.



\end{document}